\documentclass[11pt]{article}

\usepackage{color}
\usepackage{amssymb}
\usepackage{authblk}
\usepackage{times}

\expandafter\let\csname equation*\endcsname\relax 
\expandafter\let\csname endequation*\endcsname\relax 

\usepackage{xspace}
\usepackage{stmaryrd}
\usepackage{times}
\usepackage{amsfonts}
\usepackage{graphicx}
\usepackage{amsmath}
\usepackage{amssymb}
\usepackage{color}
\usepackage{subfigure}
\usepackage{tensor}
\usepackage{accents}
\usepackage[colorlinks=true, citecolor=blue]{hyperref}
\usepackage{hyperref}

\def\beq{\begin{equation}}
\def\eeq{\end{equation}}
\def\bea{\begin{eqnarray}}
\def\eea{\end{eqnarray}}
\def\ben{\begin{enumerate}}
\def\een{\end{enumerate}}

\def\b{\beta}

\def\D{\Delta}

\def\k{\kappa}

\def\O{\Omega}

\def\half{{\textstyle{\frac{1}{2}}}}

\title{\huge Scrambling and the black hole atmosphere
\vspace{1cm}}
\author{\Large Ted Jacobson${}^{1}$  and Phuc Nguyen${}^{2,3}$}
\affil{
\begin{small}
\it ${}^{1}$ Maryland Center for Fundamental Physics\\
\it University of Maryland\\
\it College Park, MD 20742\\
\vskip5mm
\it ${}^{2}$ Department of Physics and Astronomy\\
\it Lehman College, City University of New York\\
\it Bronx NY 10468, USA \\
\vskip5mm
\it ${}^{3}$ Department of Mathematics and\\
\it Haifa Research Center for Theoretical Physics and Astrophysics\\
\it University of Haifa, Haifa 3498838, Israel\\
\end{small}
}

\date{}
\begin{document}

\maketitle

\begin{abstract}
We argue that the scrambling time is the same, up to a numerical factor in three or more spacetime dimensions, as the time for the atmosphere to fall across the horizon or escape, to be replaced by new atmosphere. We propose that these times agree because the physical scrambling process is part and parcel of the atmosphere refreshment process. We provide some support for this relation also in two dimensions, but the atmosphere is not as localized, so the argument is less justified. 

\end{abstract}

\section{Introduction}

The notion that black holes are fast scramblers of quantum information was motivated by the assumption of unitarity of the black hole S-matrix \cite{Hayden:2007cs}. Several distinct time scales that might be associated with black hole scrambling converged on the same value, 
\beq\label{t*}
t_* \sim \frac{\b}{2\pi}\log S
\eeq
where $\beta$ is the inverse temperature and $S$ the entropy of the black hole. These include apparent spreading of charge that falls into a black hole, a no-cloning constraint, and a chaos timescale probed holographically with Planckian shockwave collisions 
\cite{Hayden:2007cs,Sekino:2008he,susskind2011addendum,Shenker:2013pqa}. 
A notion similar to the last of these  was introduced some three decades earlier \cite{tHooft:1984kcu}, as the time after which ``a pure state description of a black hole will be very difficult". In the shockwave analysis of \cite{Shenker:2013pqa} and some later work, the timescale depends on the energy input, and is longest when that energy is minimized. Taking the minimal energy disturbance to be one thermal unit at the Hawking temperature yields again the result 
\eqref{t*}. 
For an asymptically Anti-de Sitter black hole, according to AdS/CFT duality,  black hole scrambling is dual to thermalizing interactions of the dual, strongly coupled conformal quantum field theory, and the scrambling time in one thermal 
cell of the CFT is again of the form \eqref{t*}, but now
with $S\sim N^2$, the entropy per thermal cell. 
The nature of scrambling in various quantum systems, including holographic duals of black holes, is fairly clear, and has been studied in a number of models.
However, the nature of scrambling in terms of black hole physics remains mysterious.

At the classical level, spacetime outside a large black hole is quiescent and gently curved,  and presents no reason to suspect any kind of ``scrambling" might be taking place when a disturbance falls in across the horizon. However, the black hole is a deformed quantum field vacuum, so there is in fact much quantum activity around a black hole. Indeed, the tidal 
acceleration acting on these vacuum fluctuations produces the Hawking radiation.
Previous authors have suggested that collisions of infalling quanta with outgoing quantum field fluctuations 
are responsible for scrambling.
This mechanism was first discussed in \cite{tHooft:1984kcu}, and the idea was further explored in the setting of AdS/CFT \cite{Shenker:2013pqa}. It is an elusive notion, since the outgoing fluctuations are in their ground state. That is, infalling quanta encounter nothing but vacuum outside the black hole, so there seems to be nothing with which they might interact.  
Here we  propose a  picture of the nature of black hole scrambling that involves the outgoing quantum field fluctuations in a different way. 

The near horizon quantum vacuum is a thermal ``atmosphere". Most of the
entropy in this atmosphere comes from outgoing modes with high transverse momentum, which quickly reflect from the effective potential and fall into the black hole, while modes of sufficiently high frequency and low angular momentum escape. 
The thermal atmosphere of a black hole must therefore be continually resupplied. 
On a fixed classical spacetime background, with Lorentz invariant quantum fields, there would be no UV cutoff, and the atmosphere would be resupplied from a transPlanckian reservoir of modes
just outside the horizon. But 
that picture neglects the gravitational effect of the vacuum fluctuations, and must be
far from the truth.  In particular, the finiteness of black hole entropy implies that no such reservoir exists.
In the local vacuum state the outside field modes are entangled with partners inside the horizon, so a transPlanckian reservoir would carry an infinite entanglement entropy, exceeding the finite Bekenstein-Hawking entropy  $S_{\rm BH}$ of the black hole. 
Evidently quantum gravity somehow cuts off this would-be divergent entropy contribution,\footnote{For discussions of the possible role of quantum horizon fluctuations in this cutoff see \cite{York:1983zb,Frolov:1993ym,Jacobson:2012yt}.}
and it follows that 
the resupply mechanism for the atmosphere
must also involve quantum gravity. 

Our proposal is that scrambling is part of the atmosphere replacement mechanism. In particular, we provide evidence that the scrambling time and 
the atmosphere refresh time agree up to a numerical factor,
if the cutoff is chosen so that the atmosphere entropy matches the black hole entropy
minus the extremal value for the same charge.
The refresh time shares aspects of some other definitions of the scrambling time,  but it is conceptually distinct. 

This interpretation of scrambling provides some insight into the causality puzzle posed by the rapidity of the scrambling \cite{Hayden:2007cs}. Hayden and Preskill noted that,
when viewed as thermalization on the ``stretched horizon", the scrambling process is superluminal in the transverse dimensions.
If the stretched horizon were an ordinary thermal system, no local causal dynamics could scramble that quickly. But it is {\it not} an ordinary thermal system: 
the atmosphere continually falls into the black hole or escapes from the near horizon region,  and is replaced by ``fresh" atmosphere. 
That replacement is
itself a mysterious process. How do outgoing modes emerge from the near horizon region, in the absence of a transPlanckian reservoir to supply them \cite{Jacobson:1996zs}? It seems that the process must be nonlocal, because locally
the horizon is an unremarkable place in spacetime. Together with the agreement of the refresh time and scrambling times,
this inherent nonlocality strongly suggests
that the mechanism of scrambling is intimately related to the atmosphere resupply mechanism.

\section{Atmosphere model}

To estimate the refresh time, 
we model the atmosphere as a gas of massless field modes,
and use the ray optics approximation to evaluate the  
mode propagation.
This crude model captures the consequences of radial causality, 
which we presume are all that really matters.
While most of the entropy resides in outgoing modes that fall back across the horizon more quickly, the longest lived modes are those of the smallest angular momentum and highest frequency, some of which escape from the near horizon region. Those are also the modes that are the most relevant for the Hawking radiation. We define the refresh time as the time for those modes to escape from the ``Rindlersphere" i.e., the region where the Rindler (flat spacetime) approximation is valid.
More precisely,  for static, spherical black holes, we define the Rindlersphere as the region
where the norm of the  Killing vector is approximately proportional to the proper radial distance (normal to the Killing vector) from the horizon. 

While we lack a sharp reason for using the top of the Rindlersphere to demarcate the ``tropopause" of the black hole atmosphere, it is a natural choice because, for an evaporating black hole, the Rindlersphere is the region that is in equilibrium. 
As for the ``bottom" of the atmosphere, as discussed above,
quantum gravity presumably imposes a cutoff 
on modes close to the horizon.
If the atmosphere entropy contributed by a single species
is to be of order the Bekenstein-Hawking entropy $S_{\rm BH}=A/4l_P^2$,
we should exclude modes localized 
closer to the horizon than the Planck length $l_P$, 
as measured on a spatial hypersurface orthogonal to the horizon generating Killing vector \cite{Sorkin:2014kta,tHooft:1984kcu}.
(We consider here only spherical black holes.)  

For charged black holes we set the lower cutoff so as to 
match only the portion of the black hole
entropy above the extremal value for the same charge. We make this choice in order to match the 
scrambling time found by shockwave analytics in \cite{Leichenauer:2015nxa},
but it is not entirely unmotivated.
The extremal portion of the entropy resides in a ground state degeneracy. 
It is a different kind of entropy, and is not carried by an atmosphere that is continually refreshed. (See below for more discussion of this point.)

\section{Atmosphere refresh time}

We define the refresh time as the Killing time for a radial null geodesic to climb from the bottom to the top of the Rindlersphere.  This is well-defined in spherical symmetry if the Killing time coordinate is chosen so its level sets are orthogonal to the Killing vector.\footnote{Alternatively, we could define the refresh time using a round trip that reflects at a centrifugal barrier (or an imaginary mirror) at the top of the Rindlersphere, and falls back to the initial Killing orbit. The   
elapsed Killing time along that orbit 
does not depend on which Killing time coordinate is employed, and is just twice the time defined above.}
The spherically symmetric, static 
line element then takes the form
\beq
ds^2=-N^2 dt^2 + dy^2 + r^2 d\O^2,
\eeq
where $N=N(y)$ and $r=r(y)$. 
The coordinate $y$ is the proper radial distance orthogonal
to the Killing vector $\partial_t$, whose norm is $N$. For a black hole we set $y=0$ at the horizon, so in
the Rindlersphere we have $N(y)\approx\k y$, where $\k$ is the surface gravity.

A radial light ray satisfies $dt = dy/N$, so as it propagates from the bottom to the top of the Rindlersphere 
the elapsed Killing time is 
\beq\label{Dt}
\D t = \int_{y_b}^{y_t} \frac{dy}{N}\approx \k^{-1}\log(y_t/y_b),
\eeq
where $y_b$ and $y_t$ are the radial distances from the horizon to the bottom and top of the Rindlersphere.
For a Schwarzschild black hole in four spacetime dimensions, 
the lower cutoff is the Planck length, $y_b\sim l_P$, and
we find $y_t\sim r_+$, so the log in \eqref{Dt} is
$\sim \log (r_+/l_P)\sim \half\log S_{\rm BH}$. The refresh time
\eqref{Dt} therefore agrees with \eqref{t*}, up to the prefactor $\half$, since $\k = 2\pi T_{\rm H}/\hbar= 2\pi/\beta$.

To locate the top of the Rindlersphere
we consider the Taylor expansion of the lapse $N(y)=\k y + \dots$.
The Rindlersphere ends where the higher order terms compete with the linear one.
The even terms in $y$ 
vanish for the static metrics we are considering, because regularity at the horizon implies that 
$N^2$ must be even under $y\rightarrow -y$ reflection.
Assuming that $(N_{yyy})_+$ 
($\equiv d^3N/dy^3|_{y=0}$) 
is nonzero, 
the top of the Rindlersphere lies where 
$ \kappa y \sim |N_{yyy}|_+y^3$, i.e.\ at
$y_t\sim [{\k}/{|N_{yyy}|_+}]^{1/2}$.
The metrics we consider all satisfy $N = dr/dy$. A straightforward computation using this relation shows that 
$(N_{yyy})_+ =  \half\k [(N^2)_{rr}]_+$, 
so a general expression for the location of the top of the Rindlersphere is
\beq\label{yt}
y_t\sim|(N^2)_{rr}|_+^{-1/2}.
\eeq
For Schwarzschild black holes this yields $y_t\sim r_+[(D-2)(D-3)]^{-1/2}$.
For AdS-Schwarzschild black holes that are much larger than the AdS length $L$, and for the planar case, we find $y_t\sim L[(D-1)(D-4)]^{-1/2}$. For Reissner-Nordstrom (charged) black holes in $D=4$, 
we find $y_t\sim r_+|1-2r_-/r_+|^{-1/2}$, which yields
$y_t\sim r_+$ 
as long as $r_-$ is not equal 
(or very close) to $r_+/2$.
In exceptional cases where $(N_{yyy})_+$ vanishes, we find instead $y_t \sim|\k (N^2)_{rrr}|_+^{-1/4}$.
This happens in $D=4$ for
for planar Schwarzschild-AdS black holes and for Reissner-Nordstrom black holes with $r_-=r_+/2$. When using this higher order relation, we find that the dependence of $y_t$ on the metric parameters is the same as in the generic cases.

To locate the bottom of the atmosphere, $y_b$, we require that the atmosphere entropy accounts for the portion of the black hole entropy above the extremal value, as discussed above. 
The thermal entropy can be estimated as
$S \sim A/y_b^{D-2}$ \cite{Sorkin:2014kta, tHooft:1984kcu}. Setting this equal to  
the entropy above the extremal value,
$(A-A_0)/l_{\rm P}^{D-2}$, where 
$A_0$ is the extremal area for the given charge, yields 
\beq
\label{yb}
y_b =l_P [A/(A-A_0)]^{1/(D-2)}. 
\eeq
Close to extremality, this can differ significantly from $l_P$, and it can even become greater than $y_t$, the top of the Rindlersphere. However,
the thermal approximation we are employing should not be trusted too close to extremality \cite{Preskill:1991tb}. 
For a Reissner-Nordstrom black hole 
in $D=4$ spacetime dimensions, 
\eqref{yb} yields $y_b
=l_P [r_+/(r_+-r_-)]^{1/2}$.
The condition required for validity of the equilibrium treatment is 
$r_+-r_-\gtrsim l_P^2/r_+$, 
and as long as this holds, 
we have $y_b \lesssim r_+ \sim y_t$. That is, the bottom of the atmosphere lies below the top of the Rindler region.

\section{Examples}

We are now ready to evaluate the refresh time \eqref{Dt} for several examples. 

\subsection{Schwarzschild in $D$ dimensions} 

For a Schwarzschild black hole in any dimension $D=O(1)>2$ we have 
$y_t/y_b\sim r_+/l_P\sim S^{1/(D-2)}$, so \eqref{Dt} yields $\D t\sim [\k(D-2)]^{-1}\log S$. Apart from the prefactor
$(D-2)^{-1}$, this agrees with \eqref{t*}.

It is interesting to consider also the very
large $D\gg1$ limit, if we presume that there exists a consistent 
UV completion of large $D$ quantum gravity in which black holes have
finite entropy.\footnote{See also \cite{Holdt-Sorensen:2019tne} 
for an investigation of scrambling
for large $D$ black holes.}
Keeping track of just the leading order $D$
dependence, we have $y_t\sim r_+/D$. As for $y_b$, to apply 
our criterion that the atmosphere entropy matches the black hole entropy, 
we need to consider the large $D$ limit of entropy density of massless
fields in a thermal state. The temperature dependence is $T^{D-1}$, 
and for each polarization there is a leading order $D^{D/2}$ numerical 
factor (coming from the large phase space contributing to the high energy 
tail of the Planck distribution) \cite{PlanckN}.  
In addition, there are $D(D-3)/2$
graviton polarizations, but this does not change the leading order $D$ dependence,
so the leading $D$-dependence of the entropy density is $s\sim D^{D/2}T^{(D-1)}$.
Integrating this with the local temperature $T(y) \sim 1/y$  down to 
$y_b$ we obtain an atmosphere entropy $S_{\rm atm}\sim  D^{D/2} A/y_b^{D-2}$,
and equating this to the black hole entropy $S_{\rm BH}\sim A/l_P^{D-2}$
yields $(y_b/l_P)^{D-2}\sim D^{D/2}$, i.e.\ $y_b/l_P\sim D^{1/2}$.
Putting the results together we thus have $y_t/y_b\sim D^{-3/2}r_+/l_P$.
If this is not greater than unity, the thermal atmosphere lies entirely above the tropopause of the Rindlersphere, which means the calculation is inconsistent. However, also in this case the Hawking evaporation timescale is shorter than the Planck time, so the black hole cannot be treated semiclassically \cite{Holdt-Sorensen:2019tne}.
If we assume that the black hole is large enough 
that $r_+\gg D^{3/2}l_P$,
then the atmosphere refresh time \eqref{Dt} is
$\sim \k^{-1}\log(D^{-3/2}r_+/l_P) \sim (\k D)^{-1}\log S$. 
This indicates a $1/D$ suppression of the
refresh time compared to \eqref{t*}, 
no different from what we found for $D\sim O(1)$.

\subsection{Reissner-Nordstrom in $D=4$}

For the Reissner-Nordstrom case in $D=4$
we have from 
the above results 
$y_t\sim r_+$ and 
$y_b/l_P\sim [S/(S-S_0)]^{1/2}$ \eqref{yb}. 
Since also $r_+/l_P\sim S^{1/2}$, this yields
$y_t/y_b\sim (S-S_0)^{1/2}$, 
so the refresh time is $\Delta t \sim  \half \kappa^{-1} \log(S-S_0)$.
Up to the factor 1/2, this agrees with 
the scrambling time found in  Ref.~\cite{Leichenauer:2015nxa}
via shockwave analytics with the injected
energy of order the temperature. 
 This result 
 suggests that the scrambling does not involve the degrees of freedom  counted by the extremal entropy. This makes some sense, since the extremal entropy does not correspond to a thermal atmosphere that falls into the black hole, so does not need to be refreshed. Perhaps when a charged black hole relaxes to the extremal state, unitarity may not require the scrambling of all the degrees of freedom. 
 
 A different interpretation of the shortened scrambling for near-extremal charged black holes was proposed in Ref.~\cite{Brown:2018kvn}. The viewpoint
 adopted there was that the infalling energy behaves like a fundamental string, and thus exhibits an effective transverse spreading as its momentum relative to the local static frame increases during its fall through the long throat region between the Newtonian exosphere and the Rindlersphere. That paper argued that this earlier spreading decreases the time needed to complete scrambling at the stretched horizon, and  
reproduces the result of \cite{Leichenauer:2015nxa} while maintaining the hypothesis that all of the degrees of freedom counted by the black hole entropy are involved in the scrambling. 
Note that this argument assumes that the excitation falls freely into the black hole, which produces the effective transverse spread. But a particle need not fall freely. If it is lowered toward the horizon, and only dropped in from the top of the Rindlersphere, then the long throat should play no role. 

\subsection{Very large Schwarzschild-AdS in $D=O(1)>2$ dimensions}

For an AdS-Schwarzschild black hole with $r_+\gg L$, in any dimension $D=O(1)>2$, 
we have 
$y_t/y_b\sim L/l_P$, so Eq.~\eqref{Dt} yields $\D t\sim \k^{-1}\log (L/l_P)\sim[\k(D-2)]^{-1}\log N^2$,
where $N^2\sim (L/l_P)^{(D-2)}$ is the entropy per thermal cell of the dual $SU(N)$ gauge theory \cite{Maldacena:1997re}. 
Apart from the 
$1/(D-2)$ prefactor, this agrees with
the time estimated from charge spreading in \cite{Sekino:2008he}.
This time can also be inferred from the 
shockwave analytics of Ref.~\cite{Shenker:2013pqa}.\footnote{We thank Douglas Stanford for an explanation of this point.} The time found there is
\eqref{t*}, with $S$ the black hole entropy, if the (homogeneously) injected energy corresponds to one unit of entropy. To express this in terms of quantities associated with a single thermal cell, note that 
$S = N^2/(N^2/S)$. The numerator is the 
number of degrees of freedom in a thermal cell, while the denominator is the reciprocal of the number of thermal cells, which is equal to the entropy injected per cell, given that just one unit of entropy was injected globally.
If one unit of entropy were injected per thermal cell, the shockwave scrambling time would have been $\sim \k^{-1}\log N^2$, in agreement with our defnition of the atmosphere refresh time.

\subsection{de Sitter spacetime}

It was noticed in 
Ref.~\cite{susskind2011addendum} that the charge spreading timescale implies that de Sitter space, too, is a fast scrambler,
whose scrambling time is
$t_*\sim \k^{-1}\log(L/l_P)$, where $L$ is the de Sitter radius.\footnote{In 
\cite{susskind2011addendum} the string length was used rather than the Planck length, but this makes little difference to the log. Also, the prefactor ${\cal R}$ in Eq.~(2.3) of that paper is a typographical error.} 
Moreover, recent work \cite{Aalsma:2020aib,Geng:2020kxh}
has studied scrambling in de Sitter space via shockwave computations of chaos timescales.
A de Sitter horizon also has a thermal atmosphere \cite{Gibbons:1977mu}, to which we may apply our definition of the refresh time.
The top of the Rindlersphere is or order $\sim L$, so we obtain $y_t/y_b\sim L/l_P$, and thus $\D t\sim \k^{-1}\log(L/l_P)\sim (\k(D-2))^{-1}\log S$.

Despite the similarities with black hole horizons, 
the atmosphere refresh mechanism for de Sitter horizons
may differ, since it must take place within each static patch,
whereas for a black hole it could be connected to infinity.
While the refresh time according to our criterion is
the same as for black hole horizons, perhaps that criterion 
is not correct for de Sitter horizons.
It was suggested recently in Ref.~\cite{Susskind:2021esx} 
that the difference with black hole horizons 
is more profound, and leads to a ``hyperfast" scrambling 
time $\sim\k^{-1}$, with no dependence on the horizon entropy.

\subsection{Black holes in $D=2$ dimensions}

In two spacetime dimensions our previous considerations 
for the black hole geometries and our estimate of the thermal entropy of the atmosphere do not apply,
yet various dilaton gravity theories in two spacetime dimensions
have black hole solutions \cite{Grumiller:2002nm}.
A horizon cross section for such black holes is just a single point,
but they have an entropy determined by the value of the dilaton 
at the horizon, and an associated scrambling time
if the theory supports black hole dynamics and Hawking radiation.
In this subsection we briefly consider whether 
the close relation between
the scrambling time and the atmosphere refresh time 
may hold in such theories, despite the differences from the higher dimensional cases.
Our conclusion will be ``perhaps". 

The atmosphere in two spacetime dimensions is qualitatively 
different from that in higher dimensions. 
The entanglement entropy of a half-line in a gapped theory with 
correlation length $\xi$ and a UV fixed point with central charge
$c$ is $\frac{c}{6} \log(\xi/\epsilon)$, 
where $\epsilon$ provides a UV cutoff \cite{Headrick:2019eth}. 
Thus, unlike in higher dimensions, the 
entropy depends on the log of the cutoff rather than 
a power. Also, if the entangled field is gapless, there
is an IR divergence. Our criterion for 
locating the bottom of the atmosphere by setting the entanglement 
entropy equal to the black hole entropy thus appears not to be well-justified.
If we ignore this and postulate that $\epsilon$ should be identified with 
the cutoff $y_b$ at the bottom of the atmosphere, 
and if we identify the IR cutoff with the top of the atmosphere
$y_t$, the entanglement entropy
would be $\frac{c}{6} \log(y_t/y_b)$. 
If this is set equal to
the black hole entropy $S$,\footnote{For a near-extremal black hole
one would set it equal to 
the entropy $S-S_0$ that exceeds the extremal value $S_0$} 
one obtains
$\log(y_t/y_b) = \frac{6}{c}S$, so the 
atmosphere refresh time (\ref{Dt}) becomes
$\D t=\frac{6}{c}\kappa^{-1}S$. This strongly disagrees with the 
usual scrambling time $\kappa^{-1}\log S$.
However, since this calculation is not well-justified, we do
not regard the disagreement as necessarily indicating a 
failure of the atmosphere refresh time interpretation of 
scrambling. It may rather indicate that in two dimensions the nature of the atmosphere and its
entanglement entropy in an underlying fundamental theory with finite black hole entropy is qualitatively different from that
in higher dimensions.

While it is beyond the scope of this paper to 
attempt to characterize the black hole atmosphere in 2d theories, 
we shall just note here one possible alternative to the
estimate given above, making use of the UV cutoff
that is implied by the matrix model dual to 
two-dimensional string theory~\cite{
Das:1995vj,Das:1995jw,Hartnoll:2015fca}. According to the analysis of Ref.~\cite{Hartnoll:2015fca}, the 
entropy of an interval of length $L$ in the emergent linear dilaton vacuum,
in a region where the string coupling $g_s = e^{\Phi}$ is weak,
is $S=\frac13 \log(L\exp(-\Phi_{\rm tw} -\Phi_1/2 -\Phi_2/2))$, where 
$\Phi$ is the dilaton, which appears in the semiclassical action 
in combination with the Ricci scalar as $e^{-2\Phi} R$, 
and the string length is set to unity. 
The subscripts ``tw,1,2" label the values at the 
``tachyon wall" and the two endpoints of the interval,
and it is assumed 
here that the interval is located 
in a region where $\Phi_{1,2}<\Phi_w\ll-1$. This formula indicates that 
the effective cutoff at the interval endpoint 1 
scales with $\Phi_1$ as $\epsilon\sim e^{\Phi_1/2}$, which is exponentially smaller than the string length.
If we set $y_b$ 
equal to this cutoff, $y_b\sim e^{\Phi_{\rm hor}/2}$,
then the refresh time (\ref{Dt}) becomes 
$\D t \sim \kappa^{-1}(-\frac12 \Phi_{\rm hor}+\log y_t)$.
If black holes existed in the emergent spacetime of the matrix model,
their entropy would be determined by the coefficient 
$e^{-2\Phi_{\rm hor}}$ of the Ricci scalar in the action,
in which case this refresh time would be
$\kappa^{-1}(\frac14\log S_{BH} + \log y_t)$.
And if the entropy term were to dominate
over the $y_t$ term, 
then up to the factor $\frac14$ this would agree with 
the scrambling time. 
However, this identification of $y_b$ with the cutoff inferred
from the matrix model entanglement entropy is admittedly 
only vaguely motivated. Moreover, that theory does not
even support black hole formation~\cite{Karczmarek:2004bw}.
We are only pointing out that 
perhaps a UV cutoff set by 
stringy effects could be essential to 
the log rather than linear
dependence of the refresh time on the black hole entropy 
in two dimensions.

A recently much studied alternative to the matrix model for emergent
2d string theory that does contain black holes 
is the SYK model, which is dual to JT gravity.
In Ref.~\cite{Maldacena:2016hyu} the chaos time scale in 
the SYK model was studied via the Lyapunov exponent in
correlation functions. It was observed there 
that, although 
``the bulk theory dual to SYK has a tower
of light fields roughly similar to a string spectrum" 
with string length of order the
AdS radius, 
the contribution of the duals 
of those fields to the Lyapunov exponent 
(and hence to the scrambling time) in the SYK model is 
suppressed relative to the ``gravity" contribution 
for near-extremal black holes.
Perhaps the nature of the black hole atmosphere can be understood
in this model, allowing its refresh time to be found. 
Stringy effects would presumably impose the UV cutoff of the
atmosphere.

\section{Conclusion}

We have shown that the black hole scrambling time is the same, 
up to a numerical factor in three or more spacetime dimensions, as 
our estimate of the time for the thermal atmosphere to be refreshed. 
We propose that these times agree because the physical scrambling process is 
part and parcel of the atmosphere refreshment process. The discrepancy of the numerical factor is worrisome, but it
may not point to a fundamental flaw 
in the proposal, since our crude 
model of the atmosphere and kinematic
criterion for its refresh time may just be
too crude to accurately capture this factor.
We provide some evidence supporting
also a relation in two spacetime dimensions 
between the refresh time and the scrambling time, but
the picture is not as clear because the atmosphere is apparently 
not as localized near the horizon. 

The atmosphere refreshment process must be nonlocal, since locally nothing distinguishes the horizon.  
The picture of atmosphere refreshment thus
goes some way toward illuminating why the scrambling is nonlocal in spacetime, and therefore how it can be so fast. We have said nothing, however, that elucidates the dynamics of atmosphere replacement. The origin of the outgoing black hole modes remains as obscure as ever \cite{Jacobson:1996zs}. 
We may hope that, by having linked it to scrambling, perhaps some light may be shined on it.

\section*{Acknowledgements}

We thank Douglas Stanford and Brian Swingle for helpful discussions.
This work was supported in part by
the National Science Foundation grants PHY-1708139 
and PHY-2012139
at UMD,  PHY-1748958 at the KITP,
and PHY-1820734 at CUNY.
PN acknowledges support from Israel Science Foundation grant 447/17 for the Israel portion and from U.S. National Science Foundation grant PHY-1820734 
at CUNY for the U.S. portion of the work.


{
\bibliographystyle{JHEP}
\bibliography{atmosphere}

\providecommand{\href}[2]{#2}\begingroup\raggedright\begin{thebibliography}{10}

\bibitem{Hayden:2007cs}
P.~Hayden and J.~Preskill, \emph{{Black holes as mirrors: Quantum information
  in random subsystems}},
  \href{https://doi.org/10.1088/1126-6708/2007/09/120}{\emph{JHEP} {\bfseries
  09} (2007) 120} [\href{https://arxiv.org/abs/0708.4025}{{\ttfamily
  0708.4025}}].

\bibitem{Sekino:2008he}
Y.~Sekino and L.~Susskind, \emph{{Fast Scramblers}},
  \href{https://doi.org/10.1088/1126-6708/2008/10/065}{\emph{JHEP} {\bfseries
  10} (2008) 065} [\href{https://arxiv.org/abs/0808.2096}{{\ttfamily
  0808.2096}}].

\bibitem{susskind2011addendum}
L.~Susskind, \emph{Addendum to fast scramblers}, {\emph{arXiv:1101.6048} (2011)
  }.

\bibitem{Shenker:2013pqa}
S.~H. Shenker and D.~Stanford, \emph{{Black holes and the butterfly effect}},
  \href{https://doi.org/10.1007/JHEP03(2014)067}{\emph{JHEP} {\bfseries 03}
  (2014) 067} [\href{https://arxiv.org/abs/1306.0622}{{\ttfamily 1306.0622}}].

\bibitem{tHooft:1984kcu}
G.~'t~Hooft, \emph{{On the Quantum Structure of a Black Hole}},
  \href{https://doi.org/10.1016/0550-3213(85)90418-3}{\emph{Nucl. Phys.}
  {\bfseries B256} (1985) 727}.

\bibitem{York:1983zb}
J.~W. York, Jr., \emph{{Dynamical Origin of Black Hole Radiance}},
  \href{https://doi.org/10.1103/PhysRevD.28.2929}{\emph{Phys. Rev. D}
  {\bfseries 28} (1983) 2929}.

\bibitem{Frolov:1993ym}
V.~P. Frolov and I.~Novikov, \emph{{Dynamical origin of the entropy of a black
  hole}}, \href{https://doi.org/10.1103/PhysRevD.48.4545}{\emph{Phys. Rev. D}
  {\bfseries 48} (1993) 4545}
  [\href{https://arxiv.org/abs/gr-qc/9309001}{{\ttfamily gr-qc/9309001}}].

\bibitem{Jacobson:2012yt}
T.~Jacobson, \emph{{Gravitation and vacuum entanglement entropy}},
  \href{https://doi.org/10.1142/S0218271812420060}{\emph{Int. J. Mod. Phys. D}
  {\bfseries 21} (2012) 1242006}
  [\href{https://arxiv.org/abs/1204.6349}{{\ttfamily 1204.6349}}].

\bibitem{Jacobson:1996zs}
T.~Jacobson, \emph{{On the origin of the outgoing black hole modes}},
  \href{https://doi.org/10.1103/PhysRevD.53.7082}{\emph{Phys. Rev.} {\bfseries
  D53} (1996) 7082} [\href{https://arxiv.org/abs/hep-th/9601064}{{\ttfamily
  hep-th/9601064}}].

\bibitem{Sorkin:2014kta}
R.~D. Sorkin, \emph{{1983 paper on entanglement entropy: "On the Entropy of the
  Vacuum outside a Horizon"}},  in \emph{{Proceedings, 10th International
  Conference on General Relativity and Gravitation: Padua, Italy, July 4-9,
  1983}}, vol.~2, pp.~734--736, 1984,
  \href{https://arxiv.org/abs/1402.3589}{{\ttfamily 1402.3589}}.

\bibitem{Leichenauer:2015nxa}
S.~Leichenauer, \emph{{Thermal Corrections to Entanglement Entropy from
  Holography}}, \href{https://doi.org/10.1007/JHEP09(2015)014}{\emph{JHEP}
  {\bfseries 09} (2015) 014}
  [\href{https://arxiv.org/abs/1502.07348}{{\ttfamily 1502.07348}}].

\bibitem{Preskill:1991tb}
J.~Preskill, P.~Schwarz, A.~D. Shapere, S.~Trivedi and F.~Wilczek,
  \emph{{Limitations on the statistical description of black holes}},
  \href{https://doi.org/10.1142/S0217732391002773}{\emph{Mod.\ Phys.\ Lett.\ A}
  {\bfseries 6} (1991) 2353}.

\bibitem{Holdt-Sorensen:2019tne}
F.~Holdt-S{\o}rensen, D.~A. McGady and N.~Wintergerst, \emph{{Black hole
  evaporation and semiclassicality at large $D$}},
  \href{https://doi.org/10.1103/PhysRevD.102.026016}{\emph{Phys. Rev. D}
  {\bfseries 102} (2020) 026016}
  [\href{https://arxiv.org/abs/1908.08083}{{\ttfamily 1908.08083}}].

\bibitem{PlanckN}
S.~AL-Jaber, \emph{Planck's spectral distribution law in n dimensions},
  \href{https://doi.org/10.1023/A:1023391424838}{\emph{International Journal of
  Theoretical Physics} {\bfseries 42} (2003) 111}.

\bibitem{Brown:2018kvn}
A.~R. Brown, H.~Gharibyan, A.~Streicher, L.~Susskind, L.~Thorlacius and
  Y.~Zhao, \emph{{Falling Toward Charged Black Holes}},
  \href{https://doi.org/10.1103/PhysRevD.98.126016}{\emph{Phys.\ Rev.\ D}
  {\bfseries 98} (2018) 126016}
  [\href{https://arxiv.org/abs/1804.04156}{{\ttfamily 1804.04156}}].

\bibitem{Maldacena:1997re}
J.~M. Maldacena, \emph{{The Large N limit of superconformal field theories and
  supergravity}}, \href{https://doi.org/10.1023/A:1026654312961}{\emph{Adv.
  Theor. Math. Phys.} {\bfseries 2} (1998) 231}
  [\href{https://arxiv.org/abs/hep-th/9711200}{{\ttfamily hep-th/9711200}}].

\bibitem{Aalsma:2020aib}
L.~Aalsma and G.~Shiu, \emph{{Chaos and complementarity in de Sitter space}},
  \href{https://doi.org/10.1007/JHEP05(2020)152}{\emph{JHEP} {\bfseries 05}
  (2020) 152} [\href{https://arxiv.org/abs/2002.01326}{{\ttfamily
  2002.01326}}].

\bibitem{Geng:2020kxh}
H.~Geng, \emph{{Non-local entanglement and fast scrambling in de-Sitter
  holography}}, \href{https://doi.org/10.1016/j.aop.2021.168402}{\emph{Annals
  Phys.} {\bfseries 426} (2021) 168402}
  [\href{https://arxiv.org/abs/2005.00021}{{\ttfamily 2005.00021}}].

\bibitem{Gibbons:1977mu}
G.~Gibbons and S.~Hawking, \emph{{Cosmological Event Horizons, Thermodynamics,
  and Particle Creation}},
  \href{https://doi.org/10.1103/PhysRevD.15.2738}{\emph{Phys.\ Rev.\ D}
  {\bfseries 15} (1977) 2738}.

\bibitem{Susskind:2021esx}
L.~Susskind, \emph{{Entanglement and Chaos in De Sitter Holography: An SYK
  Example}},  \href{https://arxiv.org/abs/2109.14104}{{\ttfamily 2109.14104}}.

\bibitem{Grumiller:2002nm}
D.~Grumiller, W.~Kummer and D.~V. Vassilevich, \emph{{Dilaton gravity in
  two-dimensions}},
  \href{https://doi.org/10.1016/S0370-1573(02)00267-3}{\emph{Phys. Rept.}
  {\bfseries 369} (2002) 327}
  [\href{https://arxiv.org/abs/hep-th/0204253}{{\ttfamily hep-th/0204253}}].

\bibitem{Headrick:2019eth}
M.~Headrick, \emph{{Lectures on entanglement entropy in field theory and
  holography}},  \href{https://arxiv.org/abs/1907.08126}{{\ttfamily
  1907.08126}}.

\bibitem{Das:1995vj}
S.~R. Das, \emph{{Geometric entropy of nonrelativistic fermions and
  two-dimensional strings}},
  \href{https://doi.org/10.1103/PhysRevD.51.6901}{\emph{Phys. Rev. D}
  {\bfseries 51} (1995) 6901}
  [\href{https://arxiv.org/abs/hep-th/9501090}{{\ttfamily hep-th/9501090}}].

\bibitem{Das:1995jw}
S.~R. Das, \emph{{Degrees of freedom in two-dimensional string theory}},
  \href{https://doi.org/10.1016/0920-5632(95)00640-0}{\emph{Nucl. Phys. B Proc.
  Suppl.} {\bfseries 45BC} (1996) 224}
  [\href{https://arxiv.org/abs/hep-th/9511214}{{\ttfamily hep-th/9511214}}].

\bibitem{Hartnoll:2015fca}
S.~A. Hartnoll and E.~Mazenc, \emph{{Entanglement entropy in two dimensional
  string theory}},
  \href{https://doi.org/10.1103/PhysRevLett.115.121602}{\emph{Phys. Rev. Lett.}
  {\bfseries 115} (2015) 121602}
  [\href{https://arxiv.org/abs/1504.07985}{{\ttfamily 1504.07985}}].

\bibitem{Karczmarek:2004bw}
J.~L. Karczmarek, J.~M. Maldacena and A.~Strominger, \emph{{Black hole
  non-formation in the matrix model}},
  \href{https://doi.org/10.1088/1126-6708/2006/01/039}{\emph{JHEP} {\bfseries
  01} (2006) 039} [\href{https://arxiv.org/abs/hep-th/0411174}{{\ttfamily
  hep-th/0411174}}].

\bibitem{Maldacena:2016hyu}
J.~Maldacena and D.~Stanford, \emph{{Remarks on the Sachdev-Ye-Kitaev model}},
  \href{https://doi.org/10.1103/PhysRevD.94.106002}{\emph{Phys. Rev. D}
  {\bfseries 94} (2016) 106002}
  [\href{https://arxiv.org/abs/1604.07818}{{\ttfamily 1604.07818}}].

\end{thebibliography}\endgroup
}
\end{document}